\documentclass[]{aa}
\usepackage{graphicx}
\usepackage{amssymb,amsmath}
\usepackage{threeparttable}
\usepackage{txfonts}
\usepackage{float}
\usepackage{subfigure}
\usepackage{tabularx}
\usepackage{epsfig}
\usepackage[export]{adjustbox}

\def\kms {\rm $km~s^{-1}$}
\def\kmsmpc {\rm{$km~s^{-1}~Mpc^{-1}$~}}

\def\hbeta{H$\beta$}
\def\halfa{H$\alpha$}
\newcommand{\Msolar}{M$_{\odot}$}                           

\def\lsigma{$L-\sigma$}

\usepackage[normalem]{ulem}

\begin{document}

\title{Direct measurement of lensing amplification in Abell S1063 using a strongly lensed  high redshift HII Galaxy}

\author{Roberto Terlevich\inst{1,2}, Jorge Melnick\inst{3,4}, Elena~Terlevich\inst{1},  Ricardo~Ch{\'a}vez\inst{5.6}, Eduardo Telles\inst{4}, Fabio~Bresolin\inst{7}, Manolis~Plionis\inst{8,9}, Spyros~Basilakos\inst{10}, David Fern\'andez Arenas\inst{1}, Ana Luisa Gonz\'alez Mor\'an\inst{1}, \'Angeles I. D\'iaz\inst{11} \and Itziar Aretxaga\inst{1}. }

\institute{Instituto Nacional de Astrof{\'i}sica {\'O}ptica y Electr{\'o}nica, AP 51 y 216, 72000, Puebla, M{\'e}xico,
\and
Institute of Astronomy, University of Cambridge, Madingley Road, Cambridge CB3 0HA, UK,
\and
European Southern Observatory, Av. Alonso de Cordova 3107, Santiago, Chile, 
\and
Observatorio Nacional, Rua Jos\'e Cristino 77, 20921-400 Rio de Janeiro, Brasil
\and
Cavendish Laboratory, University of Cambridge, 19 J. J. Thomson Ave, Cambridge CB3 0HE, UK, 
\and
Kavli Institute for Cosmology, University of Cambridge, Madingley Road, Cambridge CB3 0HA, UK, 
\and
Institute for Astronomy, University of Hawaii, 2680 Woodlawn Drive, 96822 Honolulu,HI USA
\and
Physics Dept., Aristotle Univ. of Thessaloniki, Thessaloniki 54124, Greece 
\and
National Observatory of Athens, P.Pendeli, Athens, Greece,
\and
Academy of Athens, Research Center for Astronomy \& App. Math., Soranou Efesiou 4,  Athens 11527, Greece. 
\and
Departamento de F\'isica Te\'orica, Universidad Aut\'onoma de Madrid, Cantoblanco, E-28049 Madrid, Spain
}

\offprints{Roberto Terlevich~~ \email{rjt@inaoep.mx}}
\date{}

\authorrunning{Terlevich et. al}

\titlerunning{Strongly Lensed HII Galaxy}
\label{firstpage}

\abstract{ ID11 is an actively star forming extremely compact galaxy and Ly$\alpha $ emitter  at z=3.117 that is gravitationally magnified by a factor of {\bf $\sim $17} by the  cluster of galaxies  Hubble Frontier Fields AS1063.  Its observed properties  resemble those of  low luminosity HII galaxies or Giant HII regions like 30~Doradus in the LMC.

\smallskip
Using the tight correlation correlation between the Balmer-line luminosities and the width of the emission lines  (typically $L(H\beta)-\sigma(H\beta)$)  valid for HII galaxies and Giant HII regions to estimate its total luminosity, we  are able to measure the lensing amplification of ID11.  We obtain an amplification of $23 \pm 11  $ similar within errors  to the value of  $\sim$ 17   estimated or predicted by the  best lensing models of the massive cluster Abell  S1063.

\smallskip
We also compiled  from the literature luminosities and velocity dispersions for a set of lensed compact starforming regions. There is more scatter in  the \lsigma\ correlation for these lensed systems but on the whole the results tend to support the lensing models estimates of the magnification.  

\smallskip
Our result indicates that the amplification can be  independently measured using the \lsigma\ relation in lensed Giant HII regions or HII galaxies. 
It also  supports the suggestion, even if lensing model dependent, that the \lsigma\ relation is valid for low luminosity high-z objects. Ad-hoc observations of lensed starforming systems are required to accurately determine the lensing amplification.

}

 
\maketitle

\section{Introduction}
\label{intro}

Gravitational lensing is a powerful tool to study the properties of distant low luminosity objects and to estimate the mass profiles of clusters of galaxies. However, as reviewed by \citet{Kneib2011} and discussed  by \citet{Birrer2016} each mass mapping method has its own approach that translates into uncertainties in the recovered mass maps. On top of this most if not all methods suffer from mass-sheet degeneracy leading also to non-unique estimates  of the total mass of the clusters  \citep[see figure 1 of][]{Priewe2016}.
To circumvent this, \citet{Bertin2006} and \citet{Sonnenfeld2011} explored possible methods to make {\it direct} measurements of the magnification. In particular, they proposed  using the Fundamental Plane of elliptical galaxies as a standard rod, or the Tully-Fisher relation for spiral galaxies as a standard candle. Unfortunately both methods suffer from important weaknesses that hamper their use at high redshifts, like the uncertainty regarding the possible evolution of the Fundamental Plane, or the observational challenges of applying the Tully-Fisher relation to high-z spiral galaxies. In fact any distant estimator or standard candle can be used in lensed systems to measure its magnification. Indeed SNIa have already been used to directly measure the cluster magnification \citep{Rodney2015}. 
HII galaxies have also been used; \citet{Fosbury2003} found inconsistencies between the lensing models and the intrinsic luminosity predicted by the \citet{Melnick2000}\lsigma\ correlation for two massive star forming regions in the ``Lynx arc".

Recently \citet{Caminha2015} reported the discovery of multiple images of a strong lensed star forming galaxy at z=3.117 located behind the Hubble Frontier Fields1 cluster Abell S1063 (AS1063). This galaxy named ID11 is one of the lowest luminosity Ly $\alpha $ blobs found to date. 

In a recent paper \citet{Vanzella2016} reported new high quality observations of  ID11. Spectroscopy with MUSE and X-SHOOTER on the VLT indicates that ID11 is a compact ($R_{eff}\simeq67$ pc), young (age $< 20$ Myr), low mass (M $< 10^7$ \Msolar) and dust-free galaxy, thus strongly  resembling  low redshift HII galaxies or Giant HII regions like 30~Doradus  in the LMC \citep{Terlevich1981,Terlevich1991}. 
 
A key property of HII galaxies and Giant HII regions is that they follow a tight correlation between the widths of their emission lines, and their Balmer-line (typically $H\beta$) luminosities, the \lsigma\ correlation as shown in Figure 1 \citep[see][for a detailed discussion and list of references]{Chavez2014}.
Having shown that the \lsigma\ correlation provides a powerful  cosmological probe,  we have engaged on a programme to measure the emission line widths and luminosities of a large sample of HII galaxies out to redshifts of $z\sim3$ using multi-object IR spectrographs on the VLT and Keck telescopes   
 \citep[see][]{Terlevich2015}. 

As  the \lsigma\ correlation is valid  up to  redshifts of at least $z\sim2.5$ \citep{Melnick2000,Siegel2005,Terlevich2015}, and  prompted by \citet{Vanzella2016} results  on ID11, we decided to  use the \lsigma\ relation to measure directly the lensing amplification in Abell S1063 and  simultaneously check its  validity for low luminosity high-z starforming regions.

 \section{Results}

The measurements for the relevant lines from \citet{Vanzella2016} are reproduced in Table~\ref{vanzella}, where the line widths have already been corrected for instrumental broadening. The fluxes obtained combining two of the lensed images of the galaxy have not been corrected for internal reddening, but the authors indicate that the extinction in this object is negligible. They do not quote observational errors, but judging by the S/N, the error in the H$\beta$ flux is of the order of 25\%. The errors in the [OIII] line widths are of the order of 2-3 \kms and much larger for the weaker H$\beta$. Therefore we use the [OIII] line width for this exercise.

In order to apply the \lsigma\ relation we need to correct the line-widths for thermal broadening. This correction is in general quite small for oxygen, but the detection of HeII$\lambda$1640 \AA\ in emission in ID11 indicates that the ionized gas must be extremely hot. In fact \citet{Vanzella2016} reported a metallicity of 12+log(O/H) < 7.8 and  an electron temperature of $T_e = 26500 \pm 2600$ K  from the OIII]$\lambda$1666$\AA$/[OIII]$\lambda$5007$\AA$ ratio. Although this value of the electron temperature is on the high side compared to that of most low metallicity HII regions, we have used it to compute the thermal broadening of the oxygen lines as $\sigma_{th}=3.5$ \kms . 

The velocity dispersion of ID11, obtained from the weighted average of the two [OIII] lines is FWHM = 51.4 \kms $\pm$ 2.0 \kms , and corrected for thermal broadening is

\begin{equation}
\sigma_{[0III]}=\sqrt{(51.4/2.355)^2-3.5^2}=21.5 \pm 0.9 ~km ~s^{-1}
\end{equation}
which for the present purposes is not significantly different from the observed value of 21.8~\kms.

For giant HII regions and HII galaxies the Balmer lines are observed to be systematically broader than [OIII] by about 1.4~\kms  \citep{Hippelein1986},  so a correction must be included before using the \lsigma\ relation of \citet{Chavez2014} that is defined for H$\beta$.  Thus, for ID11 we estimate that the velocity dispersion of the Balmer lines is:  $\sigma_{H\beta}=23.0$~\kms
 
The\lsigma\ relation as calibrated in \citet{Chavez2012} using 92 Giant HII regions and HII galaxies is
\begin{equation}
log L(H\beta) = 4.97\pm0.10\times log(\sigma_{H\beta}) + 33.25\pm0.15~erg~s^{-1}
\end{equation}
which for $\sigma_{H\beta}$ = $23 ~km~s^{-1}$ predicts log$L(H\beta)_{pred}$ = $40.02\pm0.22~\rm erg~s^{-1}$;   the error has been computed as:

\begin{equation}
 \epsilon_{log (Lum H\beta )}=\sqrt{(log\sigma )^2 \times \epsilon _a ^2 + a^2 \times \epsilon _{log \sigma} + \epsilon _b ^2}
\end{equation}
where $a$ is the slope and $b$ the intercept of the \lsigma\ relation and  $\epsilon _a ,  \epsilon _b$ and $\epsilon _{log \sigma}$ are the errors in the slope, the intercept and the velocity dispersion respectively.

The more recent calibration of the \lsigma\ relation by \citet{Terlevich2015} with a larger sample of HII galaxies including high redshift ones with redshifts between 0.6 and 2.33 is,
\begin{equation}
log L(H\beta) = 5.055\pm0.097\times log(\sigma_{H\beta}) + 33.11\pm0.145~erg~s^{-1}
\end{equation}
in this case we obtain log$L(H\beta)_{pred}$=$39.99\pm0.21~\rm erg~s^{-1},$ again propagating the errors with equation (3).

The observed H$\beta$ flux of ID11 from Table~\ref{vanzella} is F(H$\beta)=3.1\times10^{-18} \rm erg s^{-1} cm^{-2}$  \citep[assuming negligible extinction as reported by][]{Vanzella2016}. At a redshift of $z=3.117$ and for the cosmology adopted in \citet{Chavez2014} ($H_0=74.3; \Omega_m=0.3; \Omega_0=1$), this flux corresponds to a photometric luminosity of $L(H\beta)_{obs}=41.37  \rm ~erg ~s^{-1}$. Thus, the measured amplification is, from equation (2)

\begin{equation}
\rm \mu = 10^{(41.37\pm0.1-40.02\pm0.22)}=22\pm11~~ \rm 
\end{equation}
or, from equation (4)

\begin{equation}
\rm \mu = 10^{(41.37\pm0.1-39.99\pm0.21)}=24\pm12~~ \rm  
\end{equation}
in remarkable  agreement with the magnification of $\mu\sim17$ reported by \citet{Vanzella2016} from the cluster lensing models of \citet{Caminha2015}.

We have assumed a photometric error of 0.1~dex in the observed H$\beta$ flux, but  the uncertainty in our predicted luminosity stems basically from the dispersion of the \lsigma\ relation, so it is unlikely that more and better data will yield a more stringent test of the lensing model with just one lensed HII galaxy.

\begin{figure*}
\includegraphics[width=1.3\textwidth,center]{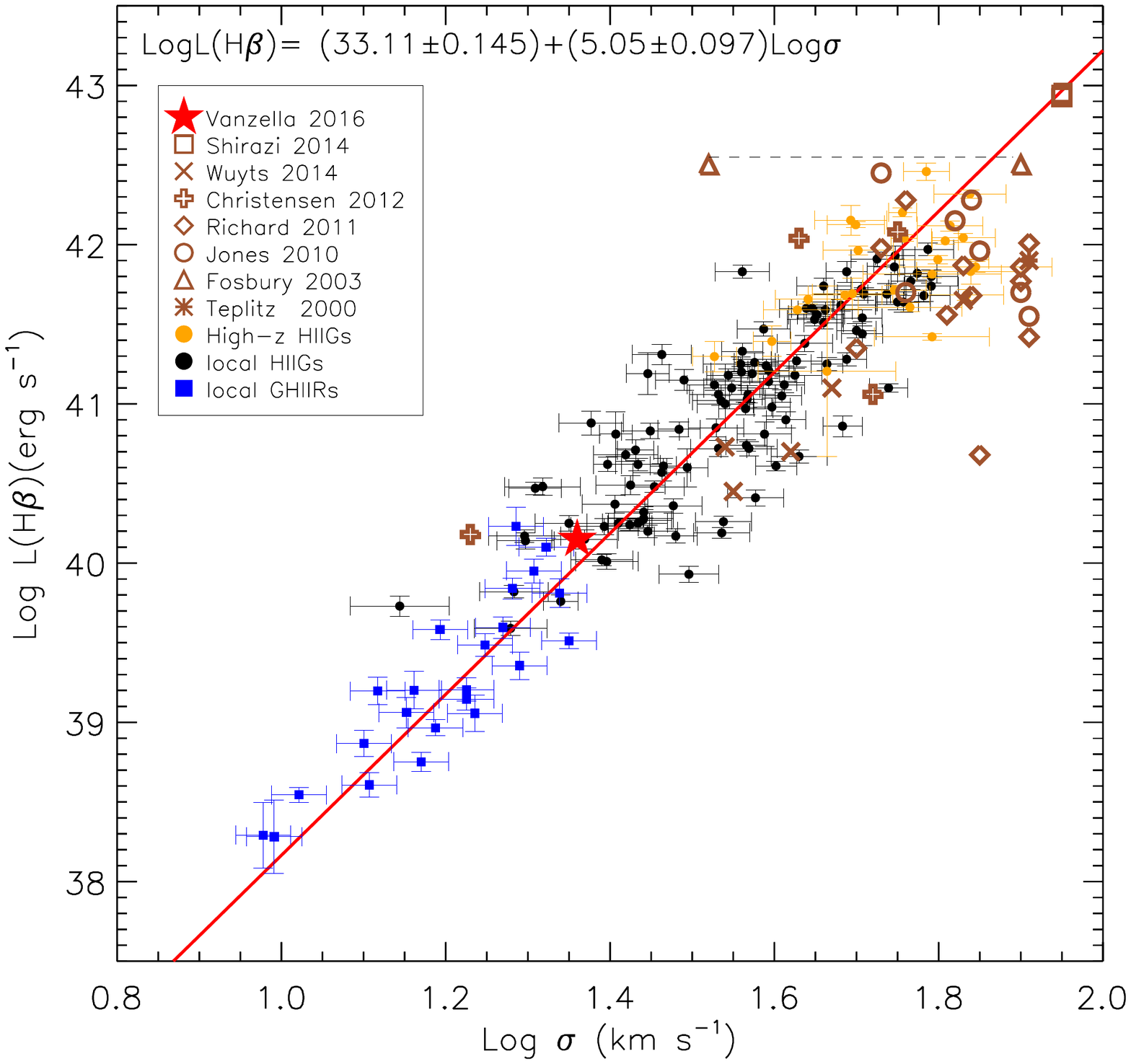} 
 \vspace*{0cm}
\caption{\small the \lsigma\ relation for HII Galaxies and Giant HII regions. Orange dots are HII galaxies with redshifts between 0.6 and 2.4.    ID11 is represented by the red  star near the centre of the plot at $\sigma_{[H\beta]}$ = 23.0$ \rm  ~km ~s^{-1}$ and  at the observed luminosity de-amplified by 0.06  $L(H\beta)_{obs}=40.15 ~\rm erg ~s^{-1}$. ID11 lies in a region of the diagram that  is populated by low luminosity HII galaxies and luminous Giant HII regions.  The open brown triangles joined by a dashed line are Fosbury's Lynx Arc data, with $\sigma$ determined from the semi-forbidden lines and from CIV (see text). Other lensed systems are plotted with a variety of symbols as listed in the inset. The brown cross near ID11 corresponds to A4.1. The  equation at the top is the fit to 131 HII Galaxies and Giant HII regions from \citet{Terlevich2015} shown by the red solid line. 
}
\label{all}
\end{figure*}

\begin{table}
   \caption{\bf Observations of ID11 from  \citet{Vanzella2016}}   
   \begin{tabular}{ c c c c c}           
 \hline\hline
Line		&	Flux 			&    S/N	&	FWHM	&	EW \\ 
		&      ($10^{-17}ergs^{-1}cm^{-2}$)	&		&	(km/s)		&	(\AA) \\ \hline
H$\beta$	&	0.31				&    4		&       ----		&	110 \\
$\rm [OIII]4959$	 &	0.90				&  12		& 	54		&	340 \\
$\rm [OIII]5007$	 &	2.35				&  33		& 	51		&	860 \\ \hline

\end{tabular}
\label{vanzella}
\end{table} 

The fact that the model demagnified luminosity (log L(H$\beta)$=40.15) agrees so well with the value expected from the \lsigma\ relation from the observed ionized gas velocity dispersion  (log L(H$\beta)$=40.02), provides strong confirmation that ID11 is either a bona-fide HII galaxy or a Giant HII region. 
This can be seen in Figure~1, where we reproduce the \lsigma\ correlation from \citet{Terlevich2015}  including nearby and high redshift HII galaxies  as well as Giant HII regions in nearby galaxies with accurate distances determined with Cepheids.

ID11 (shown by the red  star) lies in the transition region between giant HII regions and HII galaxies. Given that the deep HST images of ID11 of \citet{Vanzella2016} reveal no indication of an underlying galaxy, we concur with them in classifying  ID11 as a low-luminosity HII galaxy at z=3.117.

\section{Discussion}

As mentioned in the introduction, this is not the first time the \lsigma\ relation is used in this context.  \citet{Fosbury2003} found inconsistencies between the lensing models and the intrinsic luminosity predicted by the \citet{Melnick2000}\lsigma\ correlation on two compact and luminous star forming regions in the ``Lynx arc".   \citet{Fosbury2003} used the line profiles of two UV semi-forbidden lines, OIII]$\lambda$1666\AA\ and the  CIII]$\lambda\lambda$1907,1909\AA\ doublet  that gave values of the instrumental corrected velocity dispersions of 35 and 30 $km s^{-1}$ respectively. They also reported model corrected velocity dispersions of the absorption affected   Ly$\alpha ~\lambda$1216\AA\ and CIV$\lambda$1550\AA\  permitted lines of 150 and 80$km s^{-1}$ respectively.  In many starforming galaxies, Ly$\alpha$ is clearly asymmetric, a fact that renders its use for determining the line profile and hence velocity dispersions, very uncertain. The issue has been discussed extensively (and models constructed to reproduce it) in our own old work on Ly$\alpha$ emission \citep[e.g.][etc.]{Kunth1998,TT99,Mas-Hesse2003} and for high redshift objects \citep[][among others]{Pettini2000, Rhoads2000}. 
 On the other hand while CIV $\lambda$ 1550\AA\  is also a resonance line, resonance scattering and collisional de-excitation should affect it less than in the case of
Ly$\alpha$, because carbon is far less abundant than hydrogen and therefore has smaller optical depths -- even more so in low metallicity systems.

To  investigate further this issue we compiled  from the literature luminosities and velocity dispersions for a set of lensed compact starforming regions. The sample is listed in Table~\ref{lensed}. The lensed systems are also shown in Figure~1.

While ID11 and notably also A4.1 fall in the transition region between HII galaxies and giant HII regions the rest of the lensed systems are among the most luminous starforming systems and occupy a region of the diagram corresponding to the most luminous HII galaxies. The ``Lynx arc" result from \citet{Fosbury2003} is shown with two diferent values of the velocity dispersion, the lower value corresponding to the semi-forbidden lines and the higher value corresponding to the fit to the permitted CIV$\lambda$ 1550\AA\ line.
There is more scatter in the lensed systems but on the whole the results tend to support the lensing models estimates of the magnification apart from the ``Lynx arc" result when using the semi-forbidden lines. 

There is a crucial general point that we need to make regarding this exercise:

\noindent  The luminosity in these young bursts of star formation evolves quickly in time scales of few megayears  moving their position away from the \lsigma\ correlation to lower luminosities and smaller equivalent widths (EW) of H$\beta $ or H$\alpha $ \citep {Melnick2000, Bordalo2011}. This fact may be related to the scatter of the \citet{Richard2011} points below the \lsigma\ relation (Figure 1). To obtain and publish 
the EW of H$\beta $ or H$\alpha$ of  star forming systems is therefore highly recommended in order to allow for an evolution correction to be performed.

\section{Conclusions}

Regarding the intrinsic properties of ID11, Figure~1 implies that it is either a luminous Giant HII region like 30~Doradus in the LMC or  those found in spiral galaxies like M101, or a low luminosity HII galaxy. Given that the deep HST images of ID11 show no indication of an underlying galaxy, we confirm \citet{Vanzella2016} conclusion that ID11 is a low-luminosity HII galaxy at z=3.117.

The\lsigma\ relation spans more than three orders of magnitude in luminosity, which allows us to observe a significant number of objects at large redshifts without recourse to gravitational telescopes, hence the power of the relation as a cosmological probe \citep{Plionis2011,Terlevich2015,Chavez2016}. But, of course, only the most luminous HII galaxies can be observed at higher redshifts with the current generation of 8-10m telescopes, so gravitationally amplified objects such as ID11 provide us with the unique chance to verify that at high redshift the correlation holds also for low luminosity and possibly low metallicity HII galaxies.

Using an independent method based on the standard candle provided by the \lsigma\ relation valid for HII galaxies and Giant HII regions, we have {\bf measured} the amplification affecting the star forming system ID11 by the Abell cluster S0163 as $23\pm11$
coinciding with the  value of $\sim$17 obtained by  \citet{Vanzella2016} from the  strong lensing model of \citet{Caminha2015}.

Our result suggests that  we can use Giant HII regions in high-z lensed galaxies as instruments for studying gravitational lenses and that the combination of the \lsigma\  standard candle  plus detailed lensing models  can provide a practical method  to break the mass-sheet degeneracy in the estimates of the  total mass of clusters of galaxies.

High spectral resolution {\it ad-hoc} observations  of   H$\alpha$, H$\beta $ and [OIII]   to determine the velocity dispersion and low resolution, large aperture to determine their fluxes and EW, are needed to be able to apply this method to a larger sample of lensed starforming galaxies, but this is beyond the scope of this Letter.

\section{Acknowledgements} We  acknowledge the careful reading and constructive comments by an anonymous referee and thank Priya Natarajan for suggestions that have improved the clarity of this letter.
Roberto and Elena Terlevich  acknowledge the hospitality of the  Departmento de F{\'i}sica Te\'orica  of the Universidad Aut\'onoma de Madrid (UAM) and of the Institut d'Astrophysique de Paris (IAP),  and the financial support of the programme Study of Emission-Line Galaxies with Integral-Field Spectroscopy (SELGIFS), funded by the EU (FP7-PEOPLE-2013-IRSES-612701) within the Marie-Sklodowska-Curie Actions scheme.
Jorge Melnick acknowledges the award of a Special Visiting Researcher fellowship of the {\em Ciencia sem fronteiras} programme of the Brazilian government through their federal funding agencies - CNPq.

\begin{table*}
   \caption{\bf Parameters for lensed compact regions of star formation compiled from the literature}   
   \begin{tabular}{ l c c l l}           
 \hline\hline
Name		&  log L(H$\beta$)		&    log$\sigma$	&	$\mu$	&	REFERENCES \\ 
                      &   $ erg ~s^{-1} $  &    $km ~s^{-1}$ & &\\
\hline\hline
ID11&       40.15&  1.36&  16.7&   \cite{Vanzella2016}\\
Cl0024~*&      42.28&  1.84&  1.4&  \cite {Jones2010}~~(1)\\
MACSJ0451&    41.70&  1.90&  49.&   \cite {Jones2010}~~(1)\\
MACSJ0712&    41.55&  1.91&  28.&   \cite {Jones2010}~~(1)\\
Cl0949~*&      42.15&  1.82&  7.3&  \cite {Jones2010}~~(1)\\
Cl0949(NE)&   41.96&  1.85&  7.3&  \cite {Jones2010}~~(1)\\
Cl0949(SW)&   41.70&  1.76&  7.3&  \cite {Jones2010}~~(1)\\
MACSJ2135&    42.45&  1.73&  28.&   \cite {Jones2010}~~(1)\\
A4.1&       40.18&  1.23&  23.0&  \cite {Christensen2012}~~(2)\\
M0304&      42.08&  1.75&  42.0&  \cite {Christensen2012}~~(2)\\
M0359&      41.06&  1.72&  18.0&  \cite {Christensen2012}~~(2)\\
M2031&      42.04&  1.63&  4.2&  \cite {Christensen2012}~~(2)\\
RCSGA032727b&  40.73&  1.54&  ---&  \cite {Wuyts2014}~~(3)\\
RCSGA032727d&  40.70&  1.62&  ---&  \cite {Wuyts2014}~~(3)\\
RCSGA032727e&  41.10&  1.67&  ---&  \cite {Wuyts2014}~~(3)\\
RCSGA032727f&  40.45&  1.55&  ---& \cite { Wuyts2014}~~(3)\\
RCSGA032727g&  41.65&  1.83&  ---&  \cite {Wuyts2014}~~(3)\\
MS1512-cB58&  41.90&  1.91&  30.0&  \cite {Teplitz2000}\\
RXJ0848&     42.50&  1.52&  10.&  \cite {Fosbury2003}~~(4)\\
RXJ0848&     42.50&  1.90&  10.&  \cite {Fosbury2003}~~(5)\\
A68-C1&        40.68  & 1.85 & 2.52 & \cite {Richard2011}~~(6)\\
CEYE &         41.98  & 1.73 & 3.69 &   \cite {Richard2011}~~(6)\\
MACS0744  &    41.77 &  1.90 & 3.01 &   \cite {Richard2011}~~(6)\\
Cl0024~*  &     41.68 &  1.84 & 1.38 &   \cite {Richard2011}~~(6)\\
MACS0451 &     41.86 &  1.90 & 4.22 &   \cite {Richard2011}~~(6)\\
RXJ1053   &    41.87  & 1.83 & 4.03 &   \cite {Richard2011}~~(6)\\
A2218-Flank. &    41.35 &  1.91 & 3.60 &   \cite {Richard2011}~~(6)\\
Cl0949~*   &    41.42  & 1.91  & 2.16 &   \cite {Richard2011}~~(6)\\
A773       &   41.56  & 1.81 &  2.69 &   \cite {Richard2011}~~(6,7)\\
A2218-Mult  &  42.28  & 1.76 & 3.39  &  \cite {Richard2011}~~(6)\\
A2218-Smm    & 42.01 &  1.91 & 3.01 &   \cite {Richard2011}~~(6)\\
8OCLOCKA2   &  42.95  & 1.95 & 6.3  &  \cite {Shirazi2014}\\
8OCLOCKA3   &  42.93 &  1.95 & 4.9  &   \cite {Shirazi2014}\\

\hline

\multicolumn{5}{l}{NOTE. - Names are those listed in the corresponding reference. $\mu$ is the published magnification.}\\
\multicolumn{5}{l}{ Luminosities  recalculated from the observed fluxes for a flat universe with: }\\
\multicolumn{5}{l}{H0 = 74.3 \kmsmpc ; $\Omega_{\lambda} $= 0.70 }\\
\multicolumn{5}{l}{ }\\
\multicolumn{5}{l}{ (*) Object appears in more than one paper.}\\
\multicolumn{5}{l}{ (1) Velocity dispersion from either \halfa\ or [OIII]$\lambda$5007\AA\ . }\\
\multicolumn{5}{l}{ (2) Velocity dispersion is the weighted average of all lines from UV to optical excluding only Ly$\alpha$. }\\
\multicolumn{5}{l}{ (3) Provides de-lensed luminosities but do not list the magnifications.}\\
\multicolumn{5}{l}{ (4) Velocity dispersion  from OIII]$\lambda$1666\AA\ and  CIII]$\lambda\lambda$1907,1909\AA\  .}\\
\multicolumn{5}{l}{ (5) Velocity dispersion  from CIV$\lambda$1550\AA\ .}\\
\multicolumn{5}{l}{ (6) Velocity dispersion from either [OII]$\lambda$3727\AA , \halfa\ or [OIII]$\lambda$5007\AA ; undefined.}\\
\multicolumn{5}{l}{ (7) Used \hbeta\ flux.}\\

\end{tabular}
\label{lensed}
\end{table*}


\end{document}